\documentclass[twocolumn,showpacs,prl,superscriptaddress,floatfix]{revtex4}

\usepackage{color}
\usepackage{tabularx}
\usepackage{epsfig}
\usepackage{amsmath}
\usepackage{amssymb}
\usepackage{graphicx}
\usepackage{dcolumn}
\usepackage{bm}
\usepackage{subfigure}
\usepackage{psfrag}
\usepackage{times}

\newcommand{\LHx}{{{\rm LiHo_xY_{1-x}F_4}}}
\newcommand{\MN}{{{\rm Mn_{12}\!-\!ac}}}
\newcommand{\sss}{\scriptscriptstyle}
\newcommand{\braket}[1]{\left\langle{#1}\right\rangle}
\newcommand{\av}[1]{\left[{#1}\right]_{\sss\rm av}}
\newcommand{\aT}[1]{\braket{#1}_{\sss\rm T}}

\begin{document}

\title{Novel Disordering Mechanism in Ferromagnetic Systems with
Competing Interactions}

\author{Juan Carlos Andresen}
\affiliation {Theoretische Physik, ETH Zurich, CH-8093 Zurich,
Switzerland}

\author{Creighton K.~Thomas}
\altaffiliation{Present address: Dept.~of Materials Science and
Engineering, Northwestern University, Evanston, Illinois 60208-3108,
USA}
\affiliation {Department of Physics and Astronomy, Texas A\&M University,
College Station, Texas 77843-4242, USA}

\author{Helmut G.~Katzgraber}
\affiliation {Department of Physics and Astronomy, Texas A\&M University,
College Station, Texas 77843-4242, USA}
\address{Materials Science and Engineering Program, Texas A\&M
University, College Station, TX 77843, USA}
\affiliation {Theoretische Physik, ETH Zurich, CH-8093 Zurich, Switzerland}

\author{Moshe Schechter}
\affiliation{Department of Physics, Ben Gurion University, Beer Sheva
84105, Israel}

\date{\today}

\begin{abstract}

Ferromagnetic Ising systems with competing interactions are considered
in the presence of a random field. We find that in three space
dimensions the ferromagnetic phase is disordered by a random field which
is considerably smaller than the typical interaction strength between
the spins. This is the result of a novel disordering mechanism triggered
by an underlying spin-glass phase.  Calculations for the specific case
of the long-range dipolar $\LHx$ compound suggest that the above
mechanism is responsible for the peculiar dependence of the critical
temperature on the strength of the random field and the broadening of
the susceptibility peaks as temperature is decreased, as found in recent
experiments by Silevitch {\em et al}.~[Nature (London) {\bf 448}, 567
(2007)].  Our results thus emphasize the need to go beyond the standard
Imry-Ma argument when studying general random-field systems.

\end{abstract}


\maketitle

\paragraph*{Introduction.---}
\label{sec:intro}

The random-field Ising model (RFIM) plays a central role in the study of
disordered systems and has been applied to problems across disciplines
ranging from disordered magnets to random pinning of polymers, as well
as water seepage in porous media.

At and below the lower critical dimension $d_\ell = 2$, the
ferromagnetic (FM) phase is unstable to an infinitesimal random field
(RF) \cite{imry:75,binder:83}. At higher space dimensions the
disordering of the FM phase requires the RF strength $h$ to be of the
order of the spin-spin interaction strength $J$. Yet, the effect of the
RF on the transition between the FM and paramagnetic (PM) phases---for
systems with both short-range and dipolar interactions---has been source
of vast experimental and theoretical scrutiny
\cite{nattermann:88,belanger:98,nattermann:98}. Over the past three
decades the RFIM has been studied experimentally via dilute
antiferromagnets in a field (DAFF) \cite{fishman:79}, as both the RFIM
and the DAFF seem to share the same universality class.  More recently
it has been shown that in anisotropic dipolar magnets the RFIM can be
realized in the FM phase: By applying a transverse field to a dilute
dipolar ferromagnet, such as $\LHx$, one transforms the spatial disorder
to a longitudinal {\em effective} RF
\cite{schechter:06,tabei:06,schechter:08}. This opens the doors for
advancing our understanding of the RF problem \cite{comment:gingras}, as
well as new applications, such as tunable domain-wall pinning
\cite{silevitch:10} in magnetic materials.

Silevitch {\em et al.} recently studied the FM-to-PM transition in the
presence of RFs in $\LHx$ \cite{silevitch:07}. Remarkably, they found
that $T_c$ depends linearly on the transverse field (and thus on $h$
\cite{schechter:06,schechter:08}) and that the susceptibility peak
diminishes and broadens as temperature decreases. In $\MN$, which is a
realization of the RFIM with all FM interactions, a strong suppression
of $T_c$ as a function of $h$ was found as well \cite{wen:10}, but with
what appears to be a quite different functional dependence at small $h$.

Here we study the interplay between FM and spin-glass (SG) phases in a
dipolar Ising model with competing interactions in the presence of a RF.
We find a novel disordering mechanism of the FM phase when a RF is
applied and the system is in close proximity (e.g., via dilution) to a
SG phase.  This disordering mechanism lies between the Imry-Ma and
standard disordering mechanisms: The disordering of the FM phase occurs
at a finite RF, which is considerably smaller than the typical spin-spin
interaction, and the disordered phase [denoted henceforth as ``quasi-SG''
(QSG)] consists of not FM but glassy domains.  At $T=0$ we predict the
existence of a FM-to-QSG transition and determine for $\LHx$,
analytically and numerically, the phase boundary as a function of the Ho
concentration $x$ and RF strength $h$. At finite temperature our theory
agrees with experiments \cite{silevitch:07}, suggesting that the
existence of competing interactions and the proximity to the SG phase
dictate the broadening of the susceptibility peaks at low temperature
and the peculiar dependence of $T_c$ on $h$. Our theoretical analysis of
the SG phase follows the scaling approach of Fisher and Huse
\cite{fisher:88}---its validity supported by the agreement we find with our
numerical results. The nature of the SG phase in a RF, however, is
controversial
\cite{bhatt:85,ciria:93b,kawashima:96,billoire:03b,marinari:98d,houdayer:99,krzakala:01,takayama:04,young:04,katzgraber:05c,joerg:08a,katzgraber:09b,leuzzi:09,leuzzi:11,banos:12,larson:13},
but of no concern here.

\paragraph*{Theoretical analysis.---}
\label{sec:anal}

We first study $\LHx$ at $T=0$. For dilutions $x > x_c$ the system is
FM, whereas for $x_0 < x < x_c$ the system is a SG. Below we show
numerically that $x_c \approx 0.3$.  To date, it is unclear if $x_0 > 0$
\cite{stephen:81,snider:05}.  For $x \approx x_c$ we define the energy
per spin of the lowest FM state of the system as $f_{\rm FM}(x)$, and the
lowest energy of the SG state as $f_{\rm SG}(x)$.  Note that $f_{\rm
FM}(x)$ is the ground-state energy of the FM phase when $x > x_c$ and
$f_{\rm SG}(x)$ represents the ground-state energy of the SG phase for
$x_0 < x < x_c$.  At $x = x_c$ $f_{\rm FM}(x_c) = f_{\rm SG}(x_c)$, and
for $x \approx x_c$, to first order in $x - x_c$, $f_{\rm SG}(x)-f_{\rm
FM}(x) = \alpha(x - x_c) + \ldots$. We consider the FM phase for $x >
x_c$ in an applied RF of mean zero and standard deviation $h$. For small
$h$, the FM state in three dimensions cannot gain energy from the
field, because domain flips are not energetically favorable.  However,
for spin glasses the lower critical (Imry-Ma) dimension is infinity
\cite{fisher:88}.  In particular, in 3D the energy of the system can be
lowered by flipping domains, creating a QSG phase with a finite
correlation length.  Thus, for $x \sim x_c$ the energy of the SG state
will become lower than the energy of the FM state at a finite RF, which
is still considerably smaller than the typical spin-spin interaction
$J$.  More generally, any 3D Ising system with competing interactions
having at zero RF a FM ground state and a SG state at a somewhat higher
energy will be disordered through a transition to the QSG phase at a
finite RF whose magnitude depends on the proximity to the SG phase and
can be much smaller than $J$. Because in systems like $\LHx$ the
effective RFs are a result of quantum fluctuations
\cite{schechter:06,schechter:07}, this phase transition is a particular
case of a quantum phase transition where the quantum fluctuations of the
spins are small, involving only the spin's ground and first excited
states \cite{comment:timerev}, but where the {\em collective effect} of
all spins is strong enough to drive the transition.

\begin{figure}
\includegraphics[width=3in]{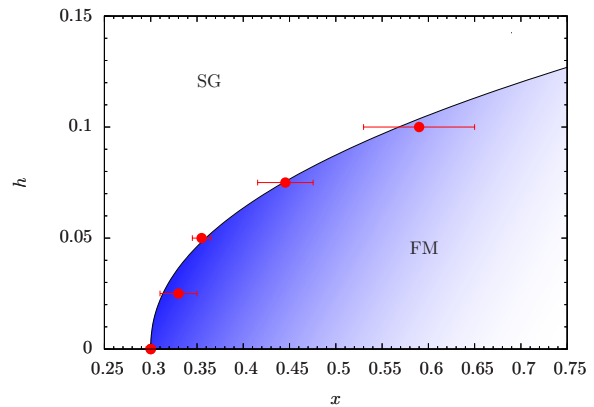}

\vspace*{-0.3cm}

\caption{(color online)
Comparison of the zero-temperature numerical and analytical
[Eq.~\eqref{hcline}] $h$-$x$ phase diagrams for the diluted dipolar
Ising model [Eq.~\eqref{HLiHo}].  $h_c(x) \sim (x - x_c)^{0.466}$, with
$\theta \approx 0.19$ \cite{hartmann:99}.  The analytical prediction
agrees well with the numerical data ($\alpha^\prime$ is a fitting
parameter).
}
\label{fig:h_x_pd}
\end{figure}

The value of the critical RF can be estimated using the short-range
Hamiltonian \cite{binder:86} in a RF $H_{\rm EA} =   - \sum_{\langle
ij\rangle} J_{ij} S_i S_j + \sum_i h_i S_i$.  $J_{ij}$ represent
nearest-neighbor Gaussian random bonds between spins $ S_i$ with zero
mean and standard deviation $J$, and $h_i$ are Gaussian RFs of average
strength $h$ \cite{comment:dipolar}.  The SG ground state is unstable to
an infinitesimal RF, creating domains of typical size
$(J/h)^{1/(3/2-\theta)}$ ($\theta \approx 0.19$) \cite{hartmann:99}. The
energy reduction per spin due to the RF is thus $f(h) =
h(J/h)^{(-3/2)/(3/2-\theta)}$.  The total energy reduction per spin is
of the same order, because the energy cost to flip domains is much
smaller. Considering now a FM system with competing interactions, e.g.,
$\LHx$ at $x>x_c$ where at $h=0$ the system is FM with $f_{\rm SG} >
f_{\rm FM}$, the critical field $h_c(x)$ can be computed from $f(h =
h_c) = f_{\rm SG} - f_{\rm FM}$, i.e., $f(h = h_c) = \alpha (x - x_c)$.
One obtains
\begin{equation}
h_c(x) = \alpha^\prime(J) (x - x_c)^{(3/2 - \theta)/(3-\theta)} \, ,
\label{hcline}
\end{equation}
where $\alpha^\prime(J) = \alpha^{(3/2 - \theta)/(3-\theta)}
J^{(3/2)/(3-\theta)}$; see Fig.~\ref{fig:h_x_pd}. For $h > h_c(x)$
there are finite domains within which glassy order persists.  The domain
size decreases with increasing field, where at $h \approx J$ the system
resembles a paramagnet. As $x \rightarrow x_c$, the disordering field
approaches zero. For large $x - x_c$ and $h \approx J$ there is a
crossover to the standard behavior where the disordering is a result of
single-spin energy minimization; i.e., the intermediate QSG regime
disappears.

We now consider finite temperatures and analyze the dependence of the FM
$T_c$ (at $x>x_c$) on the effective RF. Let us denote the lowest free
energies per spin of the FM phase (ordered for $T<T_{\rm FM}$,
disordered for $T>T_{\rm FM}$) and a competing disordered QSG phase as
$F_{\rm FM}(x,T)$ and $F_{\rm QSG}(x,T)$, respectively.  Because the
entropy of the QSG phase is dominated by regions at the boundaries
between domains \cite{fisher:88}, the main effect of the RFs is to lower
the QSG energy.  Thus, $F_{\rm QSG}(x,T) - F_{\rm FM}(x,T) = -A(T-T_{\rm
FM}) + B(x-x_c) - h/\xi_{\rm QSG}^{3/2}$ [here, for $h=0$, $F_{\rm
QSG}(x_c,T_c) = F_{\rm FM}(x_c,T_c)$]. For $h<h^* \equiv
B(x-x_c)\xi^{3/2}$ and $T=T_{\rm FM}$ we obtain $F_{\rm QSG}(x,T) >
F_{\rm FM}(x,T)$, and the transition occurs between an ordered FM phase
and a disordered PM phase dominated by FM fluctuations. However, for
$h>h^*$ we obtain $T_c(h) - T_c(0) = A^{-1} \left[ B(x-x_c) -
h/\xi^{3/2} \right]$, where the FM phase is disordered by a PM phase
dominated by fluctuations of domains having SG correlations over
distance $\xi$.  Thus, at $h=h^*$, $T_c(h)$ has a crossover from a
roughly quadratic dependence on the RF (known for a ferromagnet in a RF
\cite{ahrens:13}) to a linear dependence.  This result is supported by
our numerics.

Comparing to the experiments in Ref.~\cite{silevitch:07}, our results
are consistent with $T_c(h)$ being linear when $h \ll J$, with
deviations from linearity as $h \rightarrow 0$. Note that in
Ref.~\cite{silevitch:07} $T_c(h)$ is linear down to the lowest RFs
studied if one defines $T_c$ by the asymptotic behavior of the
susceptibility at high temperatures. However, if $T_c(h)$ is defined by
the peak position of the susceptibility, deviations from linearity are
observed at low fields \cite{comment:silevitch}.

\paragraph*{Numerical details.---}
\label{sec:nums}

$\LHx$ at low temperatures and in an external transverse magnetic field
is well described by \cite{chakraborty:04,schechter:08}
\begin{align}
\!\!{\mathcal H} \! = \! \sum_{i \ne j}
\frac{J_{ij}}{2} \epsilon_i \epsilon_j S_i S_j
+\frac{J_{\rm ex}}{2} \!\!\sum_{\langle i,j\rangle}
\epsilon_i \epsilon_j S_i S_j
+ \sum_i h_i \epsilon_i S_i \;.
\label{HLiHo}
\end{align}
Here $\epsilon_i=\left\{ 0,1\right\}$ is the occupation of the magnetic
${\rm Ho^{3+}}$ ions on a tetragonal lattice (lattice constants $a = b =
5.175${\AA} and $c = 10.75${\AA}) with four ions per unit cell
\cite{biltmo:09,tam:09}, $S_i \in \{\pm 1\}$, $h_i$ represent Gaussian
RFs with zero mean and standard deviation $h$, where $h$ is measured in
$[K]$. The magnetostatic dipolar coupling $J_{ij}$ between two ${\rm
Ho^{3+}}$ ions is given by: $J_{ij} = D(r^2_{ij} - 3z^2_{ij})/r_{ij}^5$,
where $r_{ij}= \vert {\bf r}_{i}-{\bf r}_j\vert$, ${\bf r}_i$ is the
position of the $i$th ${\rm Ho^{3+}}$ ion and $z_{ij}=\left( {\bf
r}_{i}-{\bf r}_j\right)\cdot \hat z$ is the component parallel to the
easy axis.  $D/a^3=0.214$K \cite{biltmo:07} and the nearest-neighbor
exchange is $J_{\rm ex}=0.12$K \cite{biltmo:09,comment:units}.  We use
periodic boundary conditions with Ewald sums \cite{tam:09,wang:01b}. At
zero field and no dilution, we find $T_c = 1.5316(2)$K, in agreement with
experimental results where $T_c = 1.530(5)$K \cite{mennega:84}.

\begin{figure*}[!tbh]

\includegraphics[width=0.30\textwidth]{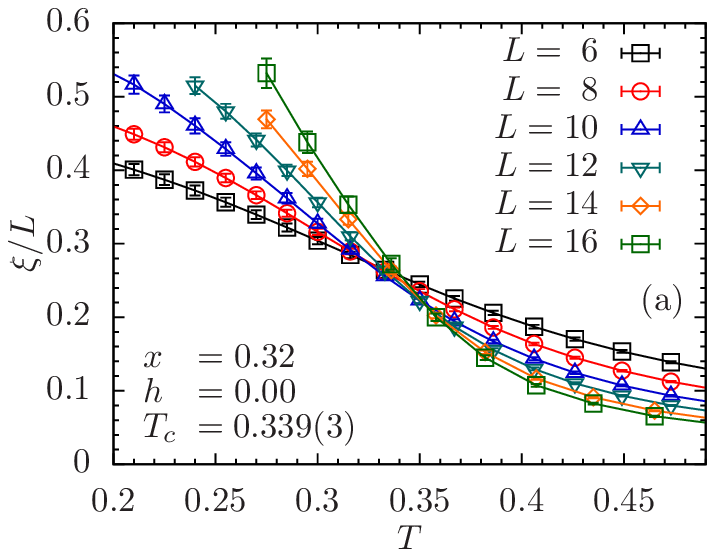}
\includegraphics[width=0.30\textwidth]{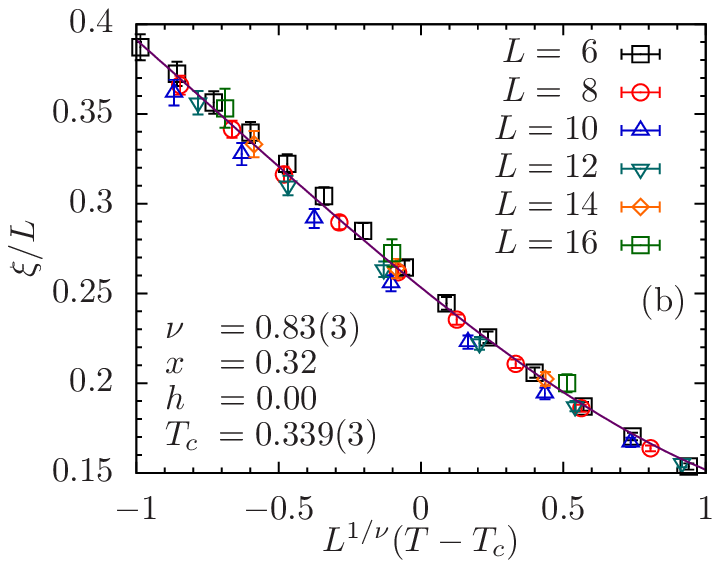}
\includegraphics[width=0.30\textwidth]{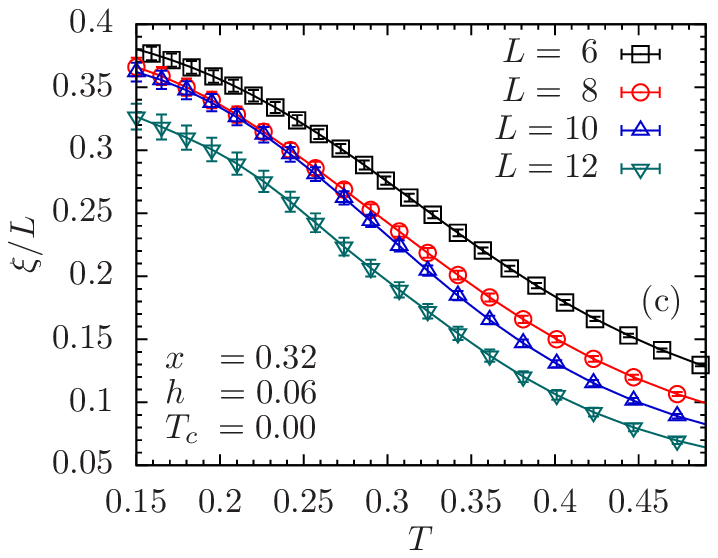}

\vspace*{-.4cm}

\caption{(color online)
Correlation length $\xi_L/L$ as a function of $T$ for $x = 0.32$. (a) $h
= 0$.  There is a clear crossing for $T_c = 0.340(3)$
\cite{comment:units} for different system sizes $L$. (b) Scaling of the
data for $h = 0$. The solid line represents the optimal scaling function
(polynomial approximation). (c) $h = 0.06$. There is no transition.
}
\label{fig:032}
\end{figure*}

\begin{figure*}[!tbh]

\includegraphics[width=0.30\textwidth]{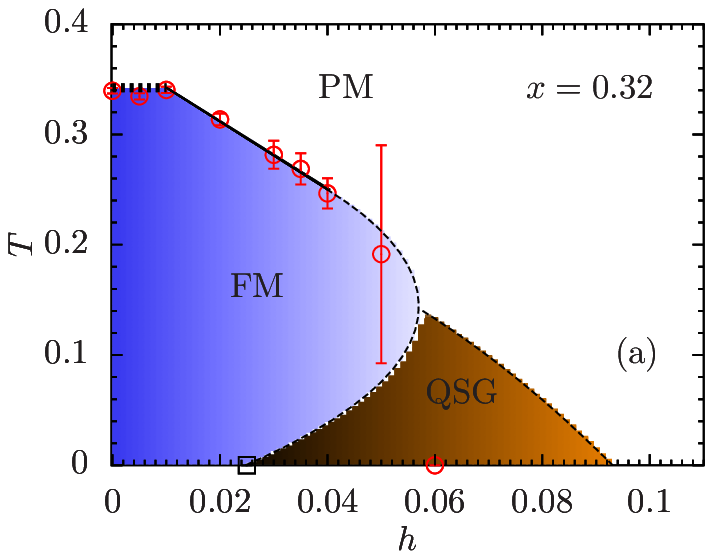}
\includegraphics[width=0.30\textwidth]{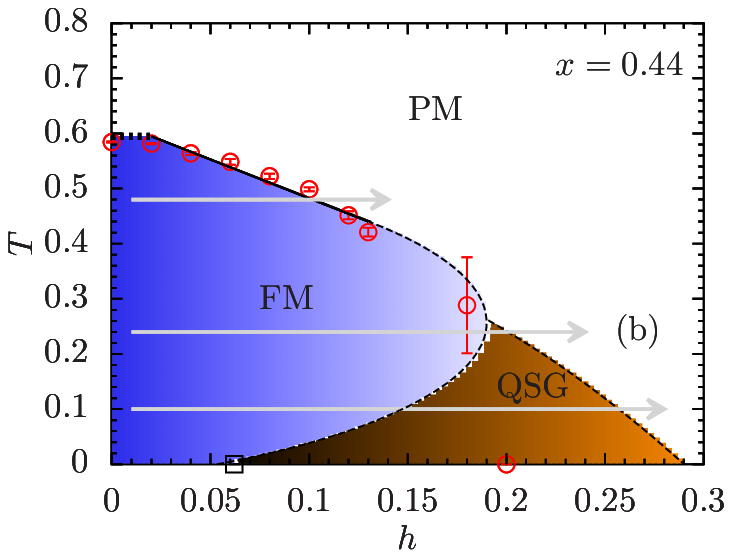}
\includegraphics[width=0.30\textwidth]{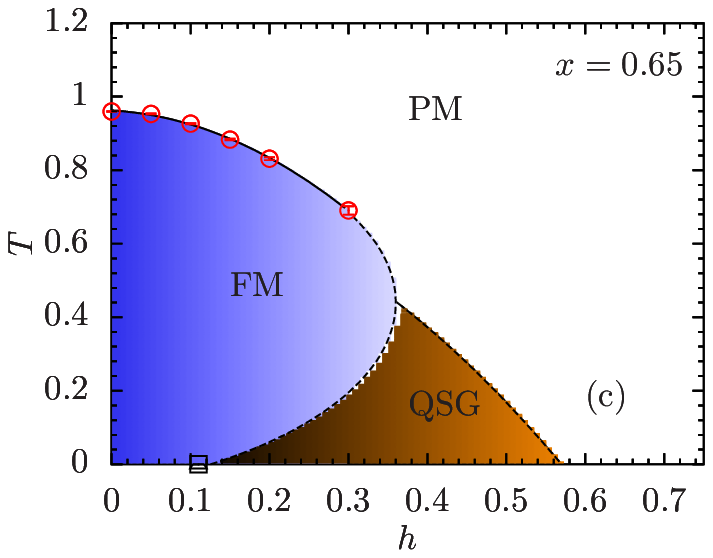}

\vspace*{-.4cm}

\caption{(color online)
$T$-$h$ phase diagram of $\LHx$ for different concentrations $x$.
Circles represent data from finite-temperature simulations, squares data
from zero-temperature simulations (see Fig.~\ref{fig:h_x_pd}). (a)
$x=0.32$.  (b) $x = 0.44$ (dilution used in the experiments of
Ref.~\cite{silevitch:07}).  Horizontal lines denote lines of constant
temperature and varying $h$, as done experimentally \cite{silevitch:07}
in susceptibility measurements.  The top line has a direct FM-PM
transition. The bottom two lines represent paths crossing the QSG phase
where a broadening of the susceptibility peak and its diminishing with
the enhancement of the RF at the crossover to the PM phase as
temperature is decreased occur.  (c) $x=0.65$. In all panels, dotted
line segments represent the conjectured phase boundary. \label{fig:pd}
}
\end{figure*}

For the zero-temperature simulations (Fig.~\ref{fig:h_x_pd}) we use
jaded extremal optimization \cite{boettcher:01,middleton:04}.  Here,
$\tau=1.6$, $1.8$ and $2$ with an aging parameter $\Gamma=0.05$ for at
least $2^{26}$ steps.  Ground states are found with high confidence for
$L\leq 10$ and $h=0$, and $L\leq 8$ with small $h\neq 0$.  The phase
boundary is identified via the Binder ratio $g = (1/2)(3 -
\av{m^4}/\av{m^2}^2)$, where $m=(1/N)\sum_i S_i$ ($N = 4xL^3$ is the
number of spins and $\av{\cdots}$ represents a disorder average.  $g
\sim \tilde{G}[ L^{1/\nu} \left(x - x_c\right)]$ is a dimensionless
function, allowing for the extraction of $x_c$ and $\nu$ for a fixed
$h$. Parameters are listed in the Supplementary Material, Table
\ref{tab:simparams0} \cite{comment:supp}.

At finite temperatures we use parallel tempering Monte Carlo
\cite{hukushima:96}. Parameters are listed in Tables
\ref{tab:simparams032}, \ref{tab:simparams044} and
\ref{tab:simparams065} in the Supplementary Material \cite{comment:supp}.  
To determine the transitions for a given $h$ and $x$ we measure
\cite{ballesteros:00}
\begin{equation}
\xi_L = \frac{1}{2\sin(k_{\rm min}/2)}\sqrt{
        \frac{\av{\aT{m^2(\mathbf{0})}}}
        {\av{\aT{m^2(\mathbf{k}_{\rm min})}}}-1
}\,,
\label{eq:correlation}
\end{equation}
where $m(\mathbf{k}) = (1/N) \sum_{i = 1} S_i
\exp(i\mathbf{k}\cdot\mathbf{R}_i)$. Here $\aT{\cdots}$ represents a
thermal average, and $\mathbf{R}_i$ is the spatial location of the spin
$S_i$, and $\mathbf{k}_{\rm min} = (2\pi/L,0,0)$.  $\xi_L/L \sim \tilde
X[L^{1/\nu}(T-T_c)]$; i.e., at the transition ($T=T_c$) the argument of
$\tilde X$ is zero (up to scaling corrections) and hence independent of
$L$ [lines of different system sizes cross [Fig.~\ref{fig:032}(a)]].
If, however, the lines do not meet, no transition occurs
[Fig.~\ref{fig:032}(c)]. To determine $T_c(h)$ we scale the data
[Fig.~\ref{fig:032}(b)]. Using a bootstrapped Levenberg-Marquardt
minimization \cite{katzgraber:06} allows us to determine the critical
parameters with statistical errors; see Table \ref{tab:numresults} of
the Supplemental Material \cite{comment:supp}. Note that for a given 
$x$ the critical exponent $\nu$ increases with $h$.

Figure \ref{fig:h_x_pd} shows the $h$-$x$ phase diagram of $\LHx$ at
zero temperature. We find excellent agreement with Eq.~\eqref{hcline},
using $\theta \approx 0.19$ \cite{hartmann:99}, i.e., $h_c(x) \sim (x -
x_c)^{0.466}$, and $\alpha^\prime$ a fitting parameter (quality of fit
$Q = 0.89$) \cite{press:95}.  Note, however, that good fits are also
possible for $0.42 \lesssim z \lesssim 0.5$ with an optimal value of $z
= 0.43(4)$ ($Q = 0.82$).

Figure \ref{fig:pd} shows finite-temperature data for different $x$.
Figure \ref{fig:pd}(a) shows $T_c(h)$ for $x=0.32$, i.e., $x - x_c =
0.02$ small. Our results at finite $T$ corroborate our theoretical model
with $h^* \approx 0.01$, where for $h<h^*$, $T_c$ is roughly independent
of $h$ (at such small fields the numerical resolution does not allow a
distinction between a constant and a parabolic dependence) and for
$h>h^*$, $T_c(h)$ decreases linearly.  The FM phase fully disorders, at
all temperatures, for $h \approx 0.055(5)$, a value slightly larger than
found from the $T = 0$ simulations, yet much smaller than the
interaction energy.  Both the disordering of the FM phase at small
fields and the linearity of $T_c(h)$ seem to persist up to $x=0.44$
[Fig.~\ref{fig:pd}(b)], the dilution used in Ref.~\cite{silevitch:07},
albeit with a less pronounced crossover at $h=h^*$. For $x=0.65$ [far
from the SG phase, Fig.~\ref{fig:pd}(c)] the behavior of $T_c(h)$
changes to a quadratic dependence for all $h<0.3$, suggesting a standard
FM-PM transition.  Critical parameters are listed in the Supplemental
Material, Table \ref{tab:numresults} \cite{comment:supp}.

\paragraph*{Phase diagram: Reentrance and experiment.---}

Our analysis for {\it zero} RF suggests that the critical concentration
$x_c=0.3$ separating the FM and SG phases depends only slightly, if at
all, on temperature. Reentrance to a SG phase is either missing or
limited to a small concentration regime, in contrast to previous
suggestions \cite{reich:90}.

At the same time, our results at finite RF at both zero and finite
temperature for all concentrations suggest that there is a range of RFs
where the system shows reentrance to a frozen QSG phase at low
temperatures \cite{comment:reentrance}.  The RF-temperature phase
diagram is shown in Fig.~\ref{fig:pd}.  Note also that the PM phase is
characterized by distinct correlations over the phase diagram: FM
fluctuating domains close to the FM phase at $h<h^*$ [dashed line in
Figs.~\ref{fig:pd}(a) and \ref{fig:pd}(b)], and SG fluctuating domains
close to the transition for $h>h^*$.  This form of the phase diagram is
strongly supported by, and provides an explanation for, the results of
Ref.~\cite{silevitch:07}, Fig.~2. For $T>0.3$K [inflection point in
Fig.~\ref{fig:pd}(b) above] there is a direct transition from the FM to
the PM [top horizontal arrow in Fig.~\ref{fig:pd}(b)], as is indeed
marked experimentally by a sharp cusp in the magnetic susceptibility.
For $T<0.3$K, however, as the transverse field (and correspondingly the
effective RF) is increased, the FM phase changes into a frozen QSG phase
and only then to the PM phase [central horizontal arrow in
Fig.~\ref{fig:pd}(b)].  Experimentally, this effect is mirrored by a
broad peak in the susceptibility at $T < 0.3$K, in good agreement with
the inflection point we find at $x=0.44$. As temperature is further
reduced, the crossover between the frozen QSG phase and the PM phase
occurs at a larger RF [bottom horizontal arrow in Fig.~\ref{fig:pd}(b)],
resulting in smaller glassy domains and the experimentally observed
diminishing peak of the susceptibility \cite{wu:93,schechter:06}.

\paragraph*{Conclusions.---}
\label{sec:conlcusions}

We propose a novel disordering mechanism for 3D ferromagnets with
competing interactions and an underlying spin-glass phase, resulting in
a disordering field which is finite, yet can be much smaller than the
interaction strength. We explain various aspects of the experiments of
Ref.~\cite{silevitch:07}, including the peculiar linear dependence of
$T_c$ on the applied transverse field and the diminishing and broadening
of the susceptibility peak with decreasing temperature. We further find
that at smaller concentrations ($x=0.32$, close to the spin-glass phase)
the reduction of $T_c$ with the RF becomes more pronounced. Our results
strongly support the notion that it is the interplay between the
competing interactions and the induced effective RF that dictate the
behavior of the $\LHx$ ferromagnet at low concentrations. Our analytical
results are generic to FM systems with competing interactions. It would
therefore be interesting to verify these results for other types of
interactions and lattice structures\cite{comment:3dvanilla}.

\begin{acknowledgments}

H.~G.~K.~acknowledges support from the SNF (Grant No.~PP002-114713) and
the NSF (Grant No.~DMR-1151387). M.S.~acknowledges support from the
Marie Curie Grant No.~PIRG-GA-2009-256313. The authors thank ETH Zurich
for CPU time on the Brutus cluster and A.~Aharony and D.~Silevitch for
useful discussions.

\end{acknowledgments}

\bibliography{refs,comments}

\begin{thebibliography}{60}
\expandafter\ifx\csname natexlab\endcsname\relax\def\natexlab#1{#1}\fi
\expandafter\ifx\csname bibnamefont\endcsname\relax
  \def\bibnamefont#1{#1}\fi
\expandafter\ifx\csname bibfnamefont\endcsname\relax
  \def\bibfnamefont#1{#1}\fi
\expandafter\ifx\csname citenamefont\endcsname\relax
  \def\citenamefont#1{#1}\fi
\expandafter\ifx\csname url\endcsname\relax
  \def\url#1{\texttt{#1}}\fi
\expandafter\ifx\csname urlprefix\endcsname\relax\def\urlprefix{URL }\fi
\providecommand{\bibinfo}[2]{#2}
\providecommand{\eprint}[2][]{\url{#2}}

\bibitem[{\citenamefont{{Imry} and {Ma}}(1975)}]{imry:75}
\bibinfo{author}{\bibfnamefont{Y.}~\bibnamefont{{Imry}}} \bibnamefont{and}
  \bibinfo{author}{\bibfnamefont{S.-K.} \bibnamefont{{Ma}}},
  \bibinfo{journal}{Phys. Rev. Lett.} \textbf{\bibinfo{volume}{35}},
  \bibinfo{pages}{1399} (\bibinfo{year}{1975}).

\bibitem[{\citenamefont{Binder}(1983)}]{binder:83}
\bibinfo{author}{\bibfnamefont{K.}~\bibnamefont{Binder}}, \bibinfo{journal}{Z.
  Phys. B - Condensed Matter} \textbf{\bibinfo{volume}{50}},
  \bibinfo{pages}{343} (\bibinfo{year}{1983}).

\bibitem[{\citenamefont{Nattermann}(1988)}]{nattermann:88}
\bibinfo{author}{\bibfnamefont{T.}~\bibnamefont{Nattermann}},
  \bibinfo{journal}{J. Phys. A} \textbf{\bibinfo{volume}{21}},
  \bibinfo{pages}{L645} (\bibinfo{year}{1988}).

\bibitem[{\citenamefont{Belanger}(1998)}]{belanger:98}
\bibinfo{author}{\bibfnamefont{D.~P.} \bibnamefont{Belanger}}, in
  \emph{\bibinfo{booktitle}{Spin Glasses and Random Fields}}, edited by
  \bibinfo{editor}{\bibfnamefont{A.~P.} \bibnamefont{Young}}
  (\bibinfo{publisher}{World Scientific}, \bibinfo{address}{Singapore},
  \bibinfo{year}{1998}), p. \bibinfo{pages}{251}.

\bibitem[{\citenamefont{Nattermann}(1998)}]{nattermann:98}
\bibinfo{author}{\bibfnamefont{T.}~\bibnamefont{Nattermann}}, in
  \emph{\bibinfo{booktitle}{Spin Glasses and Random Fields}}, edited by
  \bibinfo{editor}{\bibfnamefont{A.~P.} \bibnamefont{Young}}
  (\bibinfo{publisher}{World Scientific}, \bibinfo{address}{Singapore},
  \bibinfo{year}{1998}), p. \bibinfo{pages}{277}.

\bibitem[{\citenamefont{Fishman and Aharony}(1979)}]{fishman:79}
\bibinfo{author}{\bibfnamefont{S.}~\bibnamefont{Fishman}} \bibnamefont{and}
  \bibinfo{author}{\bibfnamefont{A.}~\bibnamefont{Aharony}},
  \bibinfo{journal}{J. Phys. C} \textbf{\bibinfo{volume}{12}},
  \bibinfo{pages}{L729} (\bibinfo{year}{1979}).

\bibitem[{\citenamefont{Schechter and Laflorencie}(2006)}]{schechter:06}
\bibinfo{author}{\bibfnamefont{M.}~\bibnamefont{Schechter}} \bibnamefont{and}
  \bibinfo{author}{\bibfnamefont{N.}~\bibnamefont{Laflorencie}},
  \bibinfo{journal}{Phys. Rev. Lett.} \textbf{\bibinfo{volume}{97}},
  \bibinfo{pages}{137204} (\bibinfo{year}{2006}).

\bibitem[{\citenamefont{Tabei et~al.}(2006)\citenamefont{Tabei, Gingras, Kao,
  Stasiak, and Fortin}}]{tabei:06}
\bibinfo{author}{\bibfnamefont{S.~M.~A.} \bibnamefont{Tabei}},
  \bibinfo{author}{\bibfnamefont{M.~J.~P.} \bibnamefont{Gingras}},
  \bibinfo{author}{\bibfnamefont{Y.-J.} \bibnamefont{Kao}},
  \bibinfo{author}{\bibfnamefont{P.}~\bibnamefont{Stasiak}}, \bibnamefont{and}
  \bibinfo{author}{\bibfnamefont{J.-Y.} \bibnamefont{Fortin}},
  \bibinfo{journal}{Phys. Rev. Lett.} \textbf{\bibinfo{volume}{97}},
  \bibinfo{pages}{237203} (\bibinfo{year}{2006}).

\bibitem[{\citenamefont{Schechter}(2008)}]{schechter:08}
\bibinfo{author}{\bibfnamefont{M.}~\bibnamefont{Schechter}},
  \bibinfo{journal}{Phys. Rev. B} \textbf{\bibinfo{volume}{77}},
  \bibinfo{pages}{020401(R)} (\bibinfo{year}{2008}).

\bibitem[{com({\natexlab{a}})}]{comment:gingras}
\bibinfo{note}{For a recent review and open problems see
  Ref.~\cite{gingras:11}.}

\bibitem[{\citenamefont{Silevitch et~al.}(2010)\citenamefont{Silevitch, Aeppli,
  and Rosenbaum}}]{silevitch:10}
\bibinfo{author}{\bibfnamefont{D.~M.} \bibnamefont{Silevitch}},
  \bibinfo{author}{\bibfnamefont{G.}~\bibnamefont{Aeppli}}, \bibnamefont{and}
  \bibinfo{author}{\bibfnamefont{T.~F.} \bibnamefont{Rosenbaum}},
  \bibinfo{journal}{Proc. Natl. Acad. Sci. U.S.A.}
  \textbf{\bibinfo{volume}{107}}, \bibinfo{pages}{2797} (\bibinfo{year}{2010}).

\bibitem[{\citenamefont{Silevitch et~al.}(2007)\citenamefont{Silevitch, Bitko,
  Brooke, Ghosh, Aeppli, and Rosenbaum}}]{silevitch:07}
\bibinfo{author}{\bibfnamefont{D.~M.} \bibnamefont{Silevitch}},
  \bibinfo{author}{\bibfnamefont{D.}~\bibnamefont{Bitko}},
  \bibinfo{author}{\bibfnamefont{J.}~\bibnamefont{Brooke}},
  \bibinfo{author}{\bibfnamefont{S.}~\bibnamefont{Ghosh}},
  \bibinfo{author}{\bibfnamefont{G.}~\bibnamefont{Aeppli}}, \bibnamefont{and}
  \bibinfo{author}{\bibfnamefont{T.~F.} \bibnamefont{Rosenbaum}},
  \bibinfo{journal}{Nature} \textbf{\bibinfo{volume}{448}},
  \bibinfo{pages}{567} (\bibinfo{year}{2007}).

\bibitem[{\citenamefont{Wen et~al.}(2010)\citenamefont{Wen, Subedi, Bo,
  Yeshurun, Sarachik, Kent, Millis, Lampropoulos, and Christou}}]{wen:10}
\bibinfo{author}{\bibfnamefont{B.}~\bibnamefont{Wen}},
  \bibinfo{author}{\bibfnamefont{P.}~\bibnamefont{Subedi}},
  \bibinfo{author}{\bibfnamefont{L.}~\bibnamefont{Bo}},
  \bibinfo{author}{\bibfnamefont{Y.}~\bibnamefont{Yeshurun}},
  \bibinfo{author}{\bibfnamefont{M.~P.} \bibnamefont{Sarachik}},
  \bibinfo{author}{\bibfnamefont{A.~D.} \bibnamefont{Kent}},
  \bibinfo{author}{\bibfnamefont{A.~J.} \bibnamefont{Millis}},
  \bibinfo{author}{\bibfnamefont{C.}~\bibnamefont{Lampropoulos}},
  \bibnamefont{and} \bibinfo{author}{\bibfnamefont{G.}~\bibnamefont{Christou}},
  \bibinfo{journal}{Phys. Rev. B} \textbf{\bibinfo{volume}{82}},
  \bibinfo{pages}{014406} (\bibinfo{year}{2010}).

\bibitem[{\citenamefont{Fisher and Huse}(1988)}]{fisher:88}
\bibinfo{author}{\bibfnamefont{D.~S.} \bibnamefont{Fisher}} \bibnamefont{and}
  \bibinfo{author}{\bibfnamefont{D.~A.} \bibnamefont{Huse}},
  \bibinfo{journal}{Phys. Rev. B} \textbf{\bibinfo{volume}{38}},
  \bibinfo{pages}{386} (\bibinfo{year}{1988}).

\bibitem[{\citenamefont{Bhatt and Young}(1985)}]{bhatt:85}
\bibinfo{author}{\bibfnamefont{R.~N.} \bibnamefont{Bhatt}} \bibnamefont{and}
  \bibinfo{author}{\bibfnamefont{A.~P.} \bibnamefont{Young}},
  \bibinfo{journal}{Phys. Rev. Lett.} \textbf{\bibinfo{volume}{54}},
  \bibinfo{pages}{924} (\bibinfo{year}{1985}).

\bibitem[{\citenamefont{Ciria et~al.}(1993)\citenamefont{Ciria, Parisi, Ritort,
  and Ruiz-Lorenzo}}]{ciria:93b}
\bibinfo{author}{\bibfnamefont{J.~C.} \bibnamefont{Ciria}},
  \bibinfo{author}{\bibfnamefont{G.}~\bibnamefont{Parisi}},
  \bibinfo{author}{\bibfnamefont{F.}~\bibnamefont{Ritort}}, \bibnamefont{and}
  \bibinfo{author}{\bibfnamefont{J.~J.} \bibnamefont{Ruiz-Lorenzo}},
  \bibinfo{journal}{J. Phys. I France} \textbf{\bibinfo{volume}{3}},
  \bibinfo{pages}{2207} (\bibinfo{year}{1993}).

\bibitem[{\citenamefont{Kawashima and Young}(1996)}]{kawashima:96}
\bibinfo{author}{\bibfnamefont{N.}~\bibnamefont{Kawashima}} \bibnamefont{and}
  \bibinfo{author}{\bibfnamefont{A.~P.} \bibnamefont{Young}},
  \bibinfo{journal}{Phys. Rev. B} \textbf{\bibinfo{volume}{53}},
  \bibinfo{pages}{R484} (\bibinfo{year}{1996}).

\bibitem[{\citenamefont{Billoire and Coluzzi}(2003)}]{billoire:03b}
\bibinfo{author}{\bibfnamefont{A.}~\bibnamefont{Billoire}} \bibnamefont{and}
  \bibinfo{author}{\bibfnamefont{B.}~\bibnamefont{Coluzzi}},
  \bibinfo{journal}{Phys. Rev. E} \textbf{\bibinfo{volume}{68}},
  \bibinfo{pages}{026131} (\bibinfo{year}{2003}).

\bibitem[{\citenamefont{Marinari et~al.}(1998)\citenamefont{Marinari, Naitza,
  and Zuliani}}]{marinari:98d}
\bibinfo{author}{\bibfnamefont{E.}~\bibnamefont{Marinari}},
  \bibinfo{author}{\bibfnamefont{C.}~\bibnamefont{Naitza}}, \bibnamefont{and}
  \bibinfo{author}{\bibfnamefont{F.}~\bibnamefont{Zuliani}},
  \bibinfo{journal}{J. Phys. A} \textbf{\bibinfo{volume}{31}},
  \bibinfo{pages}{6355} (\bibinfo{year}{1998}).

\bibitem[{\citenamefont{Houdayer and Martin}(1999)}]{houdayer:99}
\bibinfo{author}{\bibfnamefont{J.}~\bibnamefont{Houdayer}} \bibnamefont{and}
  \bibinfo{author}{\bibfnamefont{O.~C.} \bibnamefont{Martin}},
  \bibinfo{journal}{Phys. Rev. Lett.} \textbf{\bibinfo{volume}{82}},
  \bibinfo{pages}{4934} (\bibinfo{year}{1999}).

\bibitem[{\citenamefont{Krzakala et~al.}(2001)\citenamefont{Krzakala, Houdayer,
  Marinari, Martin, and Parisi}}]{krzakala:01}
\bibinfo{author}{\bibfnamefont{F.}~\bibnamefont{Krzakala}},
  \bibinfo{author}{\bibfnamefont{J.}~\bibnamefont{Houdayer}},
  \bibinfo{author}{\bibfnamefont{E.}~\bibnamefont{Marinari}},
  \bibinfo{author}{\bibfnamefont{O.~C.} \bibnamefont{Martin}},
  \bibnamefont{and} \bibinfo{author}{\bibfnamefont{G.}~\bibnamefont{Parisi}},
  \bibinfo{journal}{Phys. Rev. Lett.} \textbf{\bibinfo{volume}{87}},
  \bibinfo{pages}{197204} (\bibinfo{year}{2001}).

\bibitem[{\citenamefont{Takayama and Hukushima}(2004)}]{takayama:04}
\bibinfo{author}{\bibfnamefont{H.}~\bibnamefont{Takayama}} \bibnamefont{and}
  \bibinfo{author}{\bibfnamefont{K.}~\bibnamefont{Hukushima}},
  \bibinfo{journal}{J. Phys. Soc. Jpn.} \textbf{\bibinfo{volume}{73}},
  \bibinfo{pages}{2077} (\bibinfo{year}{2004}).

\bibitem[{\citenamefont{Young and Katzgraber}(2004)}]{young:04}
\bibinfo{author}{\bibfnamefont{A.~P.} \bibnamefont{Young}} \bibnamefont{and}
  \bibinfo{author}{\bibfnamefont{H.~G.} \bibnamefont{Katzgraber}},
  \bibinfo{journal}{Phys. Rev. Lett.} \textbf{\bibinfo{volume}{93}},
  \bibinfo{pages}{207203} (\bibinfo{year}{2004}).

\bibitem[{\citenamefont{Katzgraber and {Young}}(2005)}]{katzgraber:05c}
\bibinfo{author}{\bibfnamefont{H.~G.} \bibnamefont{Katzgraber}}
  \bibnamefont{and} \bibinfo{author}{\bibfnamefont{A.~P.}
  \bibnamefont{{Young}}}, \bibinfo{journal}{Phys. Rev. B}
  \textbf{\bibinfo{volume}{72}}, \bibinfo{pages}{184416}
  (\bibinfo{year}{2005}).

\bibitem[{\citenamefont{J{\"o}rg et~al.}(2008)\citenamefont{J{\"o}rg,
  Katzgraber, and Krzakala}}]{joerg:08a}
\bibinfo{author}{\bibfnamefont{T.}~\bibnamefont{J{\"o}rg}},
  \bibinfo{author}{\bibfnamefont{H.~G.} \bibnamefont{Katzgraber}},
  \bibnamefont{and} \bibinfo{author}{\bibfnamefont{F.}~\bibnamefont{Krzakala}},
  \bibinfo{journal}{Phys. Rev. Lett.} \textbf{\bibinfo{volume}{100}},
  \bibinfo{pages}{197202} (\bibinfo{year}{2008}).

\bibitem[{\citenamefont{Katzgraber et~al.}(2009)\citenamefont{Katzgraber,
  Larson, and Young}}]{katzgraber:09b}
\bibinfo{author}{\bibfnamefont{H.~G.} \bibnamefont{Katzgraber}},
  \bibinfo{author}{\bibfnamefont{D.}~\bibnamefont{Larson}}, \bibnamefont{and}
  \bibinfo{author}{\bibfnamefont{A.~P.} \bibnamefont{Young}},
  \bibinfo{journal}{Phys. Rev. Lett.} \textbf{\bibinfo{volume}{102}},
  \bibinfo{pages}{177205} (\bibinfo{year}{2009}).

\bibitem[{\citenamefont{{Leuzzi} et~al.}(2009)\citenamefont{{Leuzzi}, {Parisi},
  {Ricci-Tersenghi}, and {Ruiz-Lorenzo}}}]{leuzzi:09}
\bibinfo{author}{\bibfnamefont{L.}~\bibnamefont{{Leuzzi}}},
  \bibinfo{author}{\bibfnamefont{G.}~\bibnamefont{{Parisi}}},
  \bibinfo{author}{\bibfnamefont{F.}~\bibnamefont{{Ricci-Tersenghi}}},
  \bibnamefont{and} \bibinfo{author}{\bibfnamefont{J.~J.}
  \bibnamefont{{Ruiz-Lorenzo}}}, \bibinfo{journal}{Phys. Rev. Lett.}
  \textbf{\bibinfo{volume}{103}}, \bibinfo{pages}{267201}
  (\bibinfo{year}{2009}).

\bibitem[{\citenamefont{{Leuzzi} et~al.}(2011)\citenamefont{{Leuzzi}, {Parisi},
  {Ricci-Tersenghi}, and {Ruiz-Lorenzo}}}]{leuzzi:11}
\bibinfo{author}{\bibfnamefont{L.}~\bibnamefont{{Leuzzi}}},
  \bibinfo{author}{\bibfnamefont{G.}~\bibnamefont{{Parisi}}},
  \bibinfo{author}{\bibfnamefont{F.}~\bibnamefont{{Ricci-Tersenghi}}},
  \bibnamefont{and} \bibinfo{author}{\bibfnamefont{J.~J.}
  \bibnamefont{{Ruiz-Lorenzo}}}, \bibinfo{journal}{Philos. Mag.}
  \textbf{\bibinfo{volume}{91}}, \bibinfo{pages}{1917} (\bibinfo{year}{2011}).

\bibitem[{\citenamefont{{Ba{\~n}os} et~al.}(2012)\citenamefont{{Ba{\~n}os},
  {Cruz}, {Fernandez}, {Gil-Narvion}, {Gordillo-Guerrero}, {Guidetti},
  {I{\~n}iguez}, {Maiorano}, {Marinari}, {Martin-Mayor} et~al.}}]{banos:12}
\bibinfo{author}{\bibfnamefont{R.~A.} \bibnamefont{{Ba{\~n}os}}},
  \bibinfo{author}{\bibfnamefont{A.}~\bibnamefont{{Cruz}}},
  \bibinfo{author}{\bibfnamefont{L.~A.} \bibnamefont{{Fernandez}}},
  \bibinfo{author}{\bibfnamefont{J.~M.} \bibnamefont{{Gil-Narvion}}},
  \bibinfo{author}{\bibfnamefont{A.}~\bibnamefont{{Gordillo-Guerrero}}},
  \bibinfo{author}{\bibfnamefont{M.}~\bibnamefont{{Guidetti}}},
  \bibinfo{author}{\bibfnamefont{D.}~\bibnamefont{{I{\~n}iguez}}},
  \bibinfo{author}{\bibfnamefont{A.}~\bibnamefont{{Maiorano}}},
  \bibinfo{author}{\bibfnamefont{E.}~\bibnamefont{{Marinari}}},
  \bibinfo{author}{\bibfnamefont{V.}~\bibnamefont{{Martin-Mayor}}},
  \bibnamefont{et~al.}, \bibinfo{journal}{Proc. Natl. Acad. Sci. U.S.A.}
  \textbf{\bibinfo{volume}{109}}, \bibinfo{pages}{6452} (\bibinfo{year}{2012}).

\bibitem[{\citenamefont{Larson et~al.}(2013)\citenamefont{Larson, Katzgraber,
  Moore, and Young}}]{larson:13}
\bibinfo{author}{\bibfnamefont{D.}~\bibnamefont{Larson}},
  \bibinfo{author}{\bibfnamefont{H.~G.} \bibnamefont{Katzgraber}},
  \bibinfo{author}{\bibfnamefont{M.~A.} \bibnamefont{Moore}}, \bibnamefont{and}
  \bibinfo{author}{\bibfnamefont{A.~P.} \bibnamefont{Young}},
  \bibinfo{journal}{Phys. Rev. B} \textbf{\bibinfo{volume}{87}},
  \bibinfo{pages}{024414} (\bibinfo{year}{2013}).

\bibitem[{\citenamefont{Stephen and Aharony}(1981)}]{stephen:81}
\bibinfo{author}{\bibfnamefont{M.~J.} \bibnamefont{Stephen}} \bibnamefont{and}
  \bibinfo{author}{\bibfnamefont{A.}~\bibnamefont{Aharony}},
  \bibinfo{journal}{J. Phys. C} \textbf{\bibinfo{volume}{14}},
  \bibinfo{pages}{1665} (\bibinfo{year}{1981}).

\bibitem[{\citenamefont{Snider and Yu}(2005)}]{snider:05}
\bibinfo{author}{\bibfnamefont{J.}~\bibnamefont{Snider}} \bibnamefont{and}
  \bibinfo{author}{\bibfnamefont{C.~C.} \bibnamefont{Yu}},
  \bibinfo{journal}{Phys. Rev. B} \textbf{\bibinfo{volume}{72}},
  \bibinfo{pages}{214203} (\bibinfo{year}{2005}).

\bibitem[{\citenamefont{Schechter et~al.}(2007)\citenamefont{Schechter, Stamp,
  and Laflorencie}}]{schechter:07}
\bibinfo{author}{\bibfnamefont{M.}~\bibnamefont{Schechter}},
  \bibinfo{author}{\bibfnamefont{P.~C.~E.} \bibnamefont{Stamp}},
  \bibnamefont{and}
  \bibinfo{author}{\bibfnamefont{N.}~\bibnamefont{Laflorencie}},
  \bibinfo{journal}{J. Phys. Condens. Matter} \textbf{\bibinfo{volume}{19}},
  \bibinfo{pages}{145218} (\bibinfo{year}{2007}).

\bibitem[{com({\natexlab{b}})}]{comment:timerev}
\bibinfo{note}{The single-spin time-reversed state is not involved in this
  process \cite{schechter:06}.}

\bibitem[{\citenamefont{Binder and Young}(1986)}]{binder:86}
\bibinfo{author}{\bibfnamefont{K.}~\bibnamefont{Binder}} \bibnamefont{and}
  \bibinfo{author}{\bibfnamefont{A.~P.} \bibnamefont{Young}},
  \bibinfo{journal}{Rev. Mod. Phys.} \textbf{\bibinfo{volume}{58}},
  \bibinfo{pages}{801} (\bibinfo{year}{1986}).

\bibitem[{com({\natexlab{c}})}]{comment:dipolar}
\bibinfo{note}{Our results remain unchanged if dipolar interactions in a random
  system are used, and when short-range exchange interactions are included as
  is the case for ${\rm LiHo_xY_{1-x}F_4}$.}

\bibitem[{\citenamefont{Hartmann}(1999)}]{hartmann:99}
\bibinfo{author}{\bibfnamefont{A.~K.} \bibnamefont{Hartmann}},
  \bibinfo{journal}{Phys. Rev. E} \textbf{\bibinfo{volume}{59}},
  \bibinfo{pages}{84} (\bibinfo{year}{1999}).

\bibitem[{\citenamefont{Ahrens et~al.}(2013)\citenamefont{Ahrens, Xiao,
  Hartmann, and Katzgraber}}]{ahrens:13}
\bibinfo{author}{\bibfnamefont{B.}~\bibnamefont{Ahrens}},
  \bibinfo{author}{\bibfnamefont{J.}~\bibnamefont{Xiao}},
  \bibinfo{author}{\bibfnamefont{A.~K.} \bibnamefont{Hartmann}},
  \bibnamefont{and} \bibinfo{author}{\bibfnamefont{H.~G.}
  \bibnamefont{Katzgraber}} (\bibinfo{year}{2013}),
  \bibinfo{note}{(arXiv:cond-mat/1302.2480)}, \eprint{1302.2480}.

\bibitem[{com({\natexlab{d}})}]{comment:silevitch}
\bibinfo{note}{D.~M.~Silevitch (private communication).}

\bibitem[{\citenamefont{Chakraborty et~al.}(2004)\citenamefont{Chakraborty,
  Henelius, {Kj{\o}nsberg}, Sandvik, and Girvin}}]{chakraborty:04}
\bibinfo{author}{\bibfnamefont{P.~B.} \bibnamefont{Chakraborty}},
  \bibinfo{author}{\bibfnamefont{P.}~\bibnamefont{Henelius}},
  \bibinfo{author}{\bibfnamefont{H.}~\bibnamefont{{Kj{\o}nsberg}}},
  \bibinfo{author}{\bibfnamefont{A.~W.} \bibnamefont{Sandvik}},
  \bibnamefont{and} \bibinfo{author}{\bibfnamefont{S.~M.}
  \bibnamefont{Girvin}}, \bibinfo{journal}{Phys. Rev. B}
  \textbf{\bibinfo{volume}{70}}, \bibinfo{pages}{144411}
  (\bibinfo{year}{2004}).

\bibitem[{\citenamefont{Biltmo and Helenius}(2009)}]{biltmo:09}
\bibinfo{author}{\bibfnamefont{A.}~\bibnamefont{Biltmo}} \bibnamefont{and}
  \bibinfo{author}{\bibfnamefont{P.}~\bibnamefont{Helenius}},
  \bibinfo{journal}{Europhys. Lett.} \textbf{\bibinfo{volume}{87}},
  \bibinfo{pages}{27007} (\bibinfo{year}{2009}).

\bibitem[{\citenamefont{Tam and Gingras}(2009)}]{tam:09}
\bibinfo{author}{\bibfnamefont{K.-M.} \bibnamefont{Tam}} \bibnamefont{and}
  \bibinfo{author}{\bibfnamefont{M.~J.~P.} \bibnamefont{Gingras}},
  \bibinfo{journal}{Phys. Rev. Lett.} \textbf{\bibinfo{volume}{103}},
  \bibinfo{pages}{087202} (\bibinfo{year}{2009}).

\bibitem[{\citenamefont{Biltmo and Henelius}(2007)}]{biltmo:07}
\bibinfo{author}{\bibfnamefont{A.}~\bibnamefont{Biltmo}} \bibnamefont{and}
  \bibinfo{author}{\bibfnamefont{P.}~\bibnamefont{Henelius}},
  \bibinfo{journal}{Phys. Rev. B} \textbf{\bibinfo{volume}{76}},
  \bibinfo{pages}{054423} (\bibinfo{year}{2007}).

\bibitem[{com({\natexlab{e}})}]{comment:units}
\bibinfo{note}{All estimates of $T_c$ and values of $h$ are in kelvin (K).}

\bibitem[{\citenamefont{Wang and Holm}(2001)}]{wang:01b}
\bibinfo{author}{\bibfnamefont{Z.}~\bibnamefont{Wang}} \bibnamefont{and}
  \bibinfo{author}{\bibfnamefont{C.}~\bibnamefont{Holm}}, \bibinfo{journal}{J.
  Chem. Phys.} \textbf{\bibinfo{volume}{115}}, \bibinfo{pages}{6351}
  (\bibinfo{year}{2001}).

\bibitem[{\citenamefont{Mennenga et~al.}(1984)\citenamefont{Mennenga, {de
  Jongh}, and Huiskamp}}]{mennega:84}
\bibinfo{author}{\bibfnamefont{G.}~\bibnamefont{Mennenga}},
  \bibinfo{author}{\bibfnamefont{L.~J.} \bibnamefont{{de Jongh}}},
  \bibnamefont{and} \bibinfo{author}{\bibfnamefont{W.~J.}
  \bibnamefont{Huiskamp}}, \bibinfo{journal}{J. Magn. Magn. Mater.}
  \textbf{\bibinfo{volume}{44}}, \bibinfo{pages}{59} (\bibinfo{year}{1984}).

\bibitem[{\citenamefont{{Boettcher} and {Percus}}(2001)}]{boettcher:01}
\bibinfo{author}{\bibfnamefont{S.}~\bibnamefont{{Boettcher}}} \bibnamefont{and}
  \bibinfo{author}{\bibfnamefont{A.~G.} \bibnamefont{{Percus}}},
  \bibinfo{journal}{Phys. Rev. Lett.} \textbf{\bibinfo{volume}{86}},
  \bibinfo{pages}{5211} (\bibinfo{year}{2001}).

\bibitem[{\citenamefont{Middleton}(2004)}]{middleton:04}
\bibinfo{author}{\bibfnamefont{A.~A.} \bibnamefont{Middleton}},
  \bibinfo{journal}{Phys. Rev. E} \textbf{\bibinfo{volume}{69}},
  \bibinfo{pages}{055701(R)} (\bibinfo{year}{2004}).

\bibitem[{com({\natexlab{f}})}]{comment:supp}
\bibinfo{note}{See Supplemental Material for a list of all simulation
  parameters for both zero- and finite-temperature simulations, as well as a
  list of all critical parameters determined using a finite-size scaling.}

\bibitem[{\citenamefont{Hukushima and Nemoto}(1996)}]{hukushima:96}
\bibinfo{author}{\bibfnamefont{K.}~\bibnamefont{Hukushima}} \bibnamefont{and}
  \bibinfo{author}{\bibfnamefont{K.}~\bibnamefont{Nemoto}},
  \bibinfo{journal}{J. Phys. Soc. Jpn.} \textbf{\bibinfo{volume}{65}},
  \bibinfo{pages}{1604} (\bibinfo{year}{1996}).

\bibitem[{\citenamefont{Ballesteros et~al.}(2000)\citenamefont{Ballesteros,
  Cruz, Fernandez, Martin-Mayor, Pech, Ruiz-Lorenzo, Tarancon, Tellez, Ullod,
  and Ungil}}]{ballesteros:00}
\bibinfo{author}{\bibfnamefont{H.~G.} \bibnamefont{Ballesteros}},
  \bibinfo{author}{\bibfnamefont{A.}~\bibnamefont{Cruz}},
  \bibinfo{author}{\bibfnamefont{L.~A.} \bibnamefont{Fernandez}},
  \bibinfo{author}{\bibfnamefont{V.}~\bibnamefont{Martin-Mayor}},
  \bibinfo{author}{\bibfnamefont{J.}~\bibnamefont{Pech}},
  \bibinfo{author}{\bibfnamefont{J.~J.} \bibnamefont{Ruiz-Lorenzo}},
  \bibinfo{author}{\bibfnamefont{A.}~\bibnamefont{Tarancon}},
  \bibinfo{author}{\bibfnamefont{P.}~\bibnamefont{Tellez}},
  \bibinfo{author}{\bibfnamefont{C.~L.} \bibnamefont{Ullod}}, \bibnamefont{and}
  \bibinfo{author}{\bibfnamefont{C.}~\bibnamefont{Ungil}},
  \bibinfo{journal}{Phys. Rev. B} \textbf{\bibinfo{volume}{62}},
  \bibinfo{pages}{14237} (\bibinfo{year}{2000}).

\bibitem[{\citenamefont{Katzgraber et~al.}(2006)\citenamefont{Katzgraber,
  K\"orner, and Young}}]{katzgraber:06}
\bibinfo{author}{\bibfnamefont{H.~G.} \bibnamefont{Katzgraber}},
  \bibinfo{author}{\bibfnamefont{M.}~\bibnamefont{K\"orner}}, \bibnamefont{and}
  \bibinfo{author}{\bibfnamefont{A.~P.} \bibnamefont{Young}},
  \bibinfo{journal}{Phys. Rev. B} \textbf{\bibinfo{volume}{73}},
  \bibinfo{pages}{224432} (\bibinfo{year}{2006}).

\bibitem[{\citenamefont{Press et~al.}(1995)\citenamefont{Press, Teukolsky,
  Vetterling, and Flannery}}]{press:95}
\bibinfo{author}{\bibfnamefont{W.~H.} \bibnamefont{Press}},
  \bibinfo{author}{\bibfnamefont{S.~A.} \bibnamefont{Teukolsky}},
  \bibinfo{author}{\bibfnamefont{W.~T.} \bibnamefont{Vetterling}},
  \bibnamefont{and} \bibinfo{author}{\bibfnamefont{B.~P.}
  \bibnamefont{Flannery}}, \emph{\bibinfo{title}{Numerical Recipes in C}}
  (\bibinfo{publisher}{Cambridge University Press},
  \bibinfo{address}{Cambridge, England}, \bibinfo{year}{1995}).

\bibitem[{\citenamefont{Reich et~al.}(1990)\citenamefont{Reich, Ellman, Yang,
  Rosenbaum, Aeppli, and Belanger}}]{reich:90}
\bibinfo{author}{\bibfnamefont{D.~H.} \bibnamefont{Reich}},
  \bibinfo{author}{\bibfnamefont{B.}~\bibnamefont{Ellman}},
  \bibinfo{author}{\bibfnamefont{J.}~\bibnamefont{Yang}},
  \bibinfo{author}{\bibfnamefont{T.~F.} \bibnamefont{Rosenbaum}},
  \bibinfo{author}{\bibfnamefont{G.}~\bibnamefont{Aeppli}}, \bibnamefont{and}
  \bibinfo{author}{\bibfnamefont{D.~P.} \bibnamefont{Belanger}},
  \bibinfo{journal}{Phys. Rev. B} \textbf{\bibinfo{volume}{42}},
  \bibinfo{pages}{4631} (\bibinfo{year}{1990}).

\bibitem[{com({\natexlab{g}})}]{comment:reentrance}
\bibinfo{note}{Note that a reentrance to a spin-glass phase is a common effect
  found in glassy spin systems \cite{thomas:11d,ceccarelli:11a}.}

\bibitem[{\citenamefont{Wu et~al.}(1993)\citenamefont{Wu, Bitko, Rosenbaum, and
  Aeppli}}]{wu:93}
\bibinfo{author}{\bibfnamefont{W.}~\bibnamefont{Wu}},
  \bibinfo{author}{\bibfnamefont{D.}~\bibnamefont{Bitko}},
  \bibinfo{author}{\bibfnamefont{T.~F.} \bibnamefont{Rosenbaum}},
  \bibnamefont{and} \bibinfo{author}{\bibfnamefont{G.}~\bibnamefont{Aeppli}},
  \bibinfo{journal}{Phys. Rev. Lett.} \textbf{\bibinfo{volume}{71}},
  \bibinfo{pages}{1919} (\bibinfo{year}{1993}).

\bibitem[{com({\natexlab{h}})}]{comment:3dvanilla}
\bibinfo{note}{We have performed simulations of a vanilla three-dimensional
  ferromagnet with bond disorder coupled to a random field. Keeping the
  strength of the random fields fixed, as well as the ferromagnetic mean of the
  interactions, we change the width of the Gaussian-distributed bonds around
  the ferromagnetic mean. Preliminary results qualitatively agree very well
  with our proposed disordering mechanism and illustrate its generality.}

\bibitem[{\citenamefont{Gingras and Henelius}(2011)}]{gingras:11}
\bibinfo{author}{\bibfnamefont{M.~J.~P.} \bibnamefont{Gingras}}
  \bibnamefont{and} \bibinfo{author}{\bibfnamefont{P.}~\bibnamefont{Henelius}},
  \bibinfo{journal}{J. Phys.: Conf. Ser.} \textbf{\bibinfo{volume}{320}},
  \bibinfo{pages}{012001} (\bibinfo{year}{2011}).

\bibitem[{\citenamefont{Thomas and Katzgraber}(2011)}]{thomas:11d}
\bibinfo{author}{\bibfnamefont{C.~K.} \bibnamefont{Thomas}} \bibnamefont{and}
  \bibinfo{author}{\bibfnamefont{H.~G.} \bibnamefont{Katzgraber}},
  \bibinfo{journal}{Phys. Rev. E} \textbf{\bibinfo{volume}{84}},
  \bibinfo{pages}{040101(R)} (\bibinfo{year}{2011}).

\bibitem[{\citenamefont{Ceccarelli et~al.}(2011)\citenamefont{Ceccarelli,
  Pelissetto, and Vicari}}]{ceccarelli:11a}
\bibinfo{author}{\bibfnamefont{G.}~\bibnamefont{Ceccarelli}},
  \bibinfo{author}{\bibfnamefont{A.}~\bibnamefont{Pelissetto}},
  \bibnamefont{and} \bibinfo{author}{\bibfnamefont{E.}~\bibnamefont{Vicari}},
  \bibinfo{journal}{Phys. Rev. B} \textbf{\bibinfo{volume}{84}},
  \bibinfo{pages}{134202} (\bibinfo{year}{2011}).

\end{thebibliography}


\newpage

\hrule

\begin{center}
{\large {\bf Supplementary Material: Andresen {\em et al.}}}
\end{center}

\begin{table}[h]
\caption{
Simulation parameters for $T = 0$: System of size $L = 6$, $8$, and $10$,
field $h$ and dilution $x$ are studied.
$x_{\rm min}$ [$x_{\rm max}$] is the smallest [largest] concentration
studied and $\Delta x$ is the step size between
measurements. $N_{\rm sa}$ is the number of disorder realizations.
\label{tab:simparams0}}
{\scriptsize
\begin{tabular*}{\columnwidth}{@{\extracolsep{\fill}} l r r r r}
\hline
\hline
$h$     & $x_{\rm min}$ & $x_{\rm max}$ & $\Delta x$ & $N_{\rm sa}$ \\
\hline
$0.000$ & $0.280$       & $0.350$       & $0.010$    & $5000$       \\
$0.025$ & $0.275$       & $0.400$       & $0.025$    & $3000$       \\
$0.050$ & $0.300$       & $0.400$       & $0.025$    & $3000$       \\
$0.075$ & $0.300$       & $0.600$       & $0.050$    & $1500$       \\
$0.100$ & $0.400$       & $0.800$       & $0.100$    & $1500$       \\
\hline
\hline
\end{tabular*}
}
\end{table}

\begin{table}
\caption{
Simulation parameters at finite temperature and $x=0.32$ for different
fields $h$ and system sizes $L$. The equilibration/measurement times
are $2^b$ Monte Carlo sweeps. $T_{\rm min}$ [$T_{\rm max}$] is the
lowest [highest] temperature used and $N_T$ is the number of temperatures.
$N_{\rm sa}$ is the number of disorder realizations.
\label{tab:simparams032}}
{\scriptsize
\begin{tabular*}{\columnwidth}{@{\extracolsep{\fill}} l l r c r r c r}
\hline
\hline
$x$ & $h$ & $L$ & $b$  & $T_{\rm min}$ & $T_{\rm max}$ & $N_{T}$ & $N_{\rm sa}$ \\
\hline
$0.32$ & $0.000$ &  $6$ & $15$ & $0.100$ & $0.500$ & $25$ & $2000$ \\
$0.32$ & $0.000$ &  $8$ & $17$ & $0.100$ & $0.500$ & $25$ & $2000$ \\
$0.32$ & $0.000$ &  $10$ & $18$ & $0.168$ & $0.500$ & $20$ & $1000$ \\
$0.32$ & $0.000$ &  $12$ & $18$ & $0.240$ & $0.500$ & $15$ & $1000$ \\
$0.32$ & $0.000$ &  $14$ & $16$ & $0.275$ & $0.500$ & $10$ & $750$ \\
$0.32$ & $0.000$ &  $16$ & $16$ & $0.275$ & $0.500$ & $10$ & $385$ \\
\hline
$0.32$ & $0.005$ & $6$ & $12$ & $0.280$ & $0.550$ & $20$ & $2000$ \\
$0.32$ & $0.005$ & $8$ & $13$ & $0.280$ & $0.550$ & $20$ & $3500$ \\
$0.32$ & $0.005$ & $10$ & $15$ & $0.280$ & $0.550$ & $20$ & $2000$ \\
$0.32$ & $0.005$ & $12$ & $17$ & $0.280$ & $0.550$ & $20$ & $1200$ \\
$0.32$ & $0.005$ & $14$ & $16$ & $0.312$ & $0.550$ & $17$ & $1200$ \\
\hline
$0.32$ & $0.010$ & $6$ & $15$ & $0.050$ & $0.500$ & $30$ & $1500$ \\
$0.32$ & $0.010$ & $8$ & $17$ & $0.050$ & $0.500$ & $30$ & $1100$ \\
$0.32$ & $0.010$ & $10$ & $18$ & $0.230$ & $0.500$ & $15$ & $1000$ \\
$0.32$ & $0.010$ & $12$ & $19$ & $0.245$ & $0.500$ & $15$ & $850$ \\
$0.32$ & $0.010$ & $14$ & $17$ & $0.265$ & $0.500$ & $15$ & $750$ \\
\hline
$0.32$ & $0.020$ & $6$ & $15$ & $0.050$ & $0.500$ & $30$ & $1500$ \\
$0.32$ & $0.020$ & $8$ & $18$ & $0.050$ & $0.500$ & $30$ & $1000$ \\
$0.32$ & $0.020$ & $10$ & $16$ & $0.245$ & $0.500$ & $15$ & $1000$ \\
$0.32$ & $0.020$ & $12$ & $19$ & $0.245$ & $0.500$ & $15$ & $600$ \\
$0.32$ & $0.020$ & $14$ & $17$ & $0.274$ & $0.500$ & $13$ & $7500$ \\
\hline
$0.32$ & $0.030$ & $6$ & $14$ & $0.226$ & $0.450$ & $14$ & $3000$ \\
$0.32$ & $0.030$ & $8$ & $17$ & $0.212$ & $0.450$ & $15$ & $2000$ \\
$0.32$ & $0.030$ & $10$ & $17$ & $0.226$ & $0.450$ & $14$ & $2000$ \\
$0.32$ & $0.030$ & $12$ & $19$ & $0.226$ & $0.450$ & $14$ & $600$ \\
\hline
$0.32$ & $0.035$ & $6$ & $13$ & $0.218$ & $0.450$ & $20$ & $3000$ \\
$0.32$ & $0.035$ & $8$ & $14$ & $0.218$ & $0.450$ & $20$ & $2000$ \\
$0.32$ & $0.035$ & $10$ & $15$ & $0.218$ & $0.450$ & $20$ & $2000$ \\
$0.32$ & $0.035$ & $12$ & $18$ & $0.218$ & $0.450$ & $20$ & $650$ \\
\hline
$0.32$ & $0.040$ & $6$ & $14$ & $0.218$ & $0.450$ & $20$ & $3000$ \\
$0.32$ & $0.040$ & $8$ & $16$ & $0.218$ & $0.450$ & $20$ & $2800$ \\
$0.32$ & $0.040$ & $10$ & $17$ & $0.218$ & $0.450$ & $20$ & $2000$ \\
$0.32$ & $0.040$ & $12$ & $17$ & $0.218$ & $0.450$ & $14$ & $1000$ \\
\hline
$0.32$ & $0.050$ & $6$ & $13$ & $0.105$ & $0.500$ & $20$ & $1500$ \\
$0.32$ & $0.050$ & $8$ & $16$ & $0.150$ & $0.500$ & $20$ & $1000$ \\
$0.32$ & $0.050$ & $10$ & $18$ & $0.150$ & $0.500$ & $20$ & $1000$ \\
$0.32$ & $0.050$ & $12$ & $19$ & $0.150$ & $0.500$ & $20$ & $1000$ \\
\hline
$0.32$ & $0.060$ & $6$ & $12$ & $0.150$ & $0.500$ & $20$ & $2000$ \\
$0.32$ & $0.060$ & $8$ & $16$ & $0.150$ & $0.500$ & $20$ & $2000$ \\
$0.32$ & $0.060$ & $10$ & $19$ & $0.150$ & $0.500$ & $20$ & $2000$ \\
$0.32$ & $0.060$ & $12$ & $20$ & $0.150$ & $0.500$ & $20$ & $1000$ \\
\hline
\hline
\end{tabular*}
}
\end{table}

\vspace*{-0.4cm}

\begin{table}
\caption{
Simulation parameters at finite temperature and $x=0.44$. For details
see Table \ref{tab:simparams032}.
\label{tab:simparams044}}
{\scriptsize
\begin{tabular*}{\columnwidth}{@{\extracolsep{\fill}} l l r c r r c r}
\hline
\hline
$x$ & $h$ & $L$ & $b$  & $T_{\rm min}$ & $T_{\rm max}$ & $N_{T}$ & $N_{\rm sa}$ \\
\hline
\hline
$0.44$ & $0.000$ & $6$ & $10$ & $0.500$ & $1.000$ & $30$ & $1000$ \\
$0.44$ & $0.000$ & $8$ & $12$ & $0.500$ & $1.000$ & $30$ & $1300$ \\
$0.44$ & $0.000$ & $10$ & $13$ & $0.500$ & $1.000$ & $30$ & $2500$ \\
$0.44$ & $0.000$ & $12$ & $14$ & $0.500$ & $1.000$ & $30$ & $550$ \\
$0.44$ & $0.000$ & $14$ & $14$ & $0.560$ & $1.000$ & $47$ & $650$ \\
\hline
$0.44$ & $0.020$ & $6$ & $10$ & $0.525$ & $0.800$ & $30$ & $2000$ \\
$0.44$ & $0.020$ & $8$ & $12$ & $0.525$ & $0.800$ & $30$ & $2000$ \\
$0.44$ & $0.020$ & $10$ & $13$ & $0.525$ & $0.800$ & $30$ & $1000$ \\
$0.44$ & $0.020$ & $12$ & $14$ & $0.525$ & $0.800$ & $30$ & $1000$ \\
$0.44$ & $0.020$ & $14$ & $14$ & $0.550$ & $0.800$ & $30$ & $900$ \\
\hline
$0.44$ & $0.040$ & $6$ & $11$ & $0.525$ & $0.750$ & $22$ & $2000$ \\
$0.44$ & $0.040$ & $8$ & $12$ & $0.525$ & $0.750$ & $22$ & $2000$ \\
$0.44$ & $0.040$ & $10$ & $14$ & $0.540$ & $0.750$ & $20$ & $1000$ \\
$0.44$ & $0.040$ & $12$ & $15$ & $0.550$ & $0.750$ & $20$ & $1000$ \\
\hline
$0.44$ & $0.060$ & $6$ & $11$ & $0.500$ & $0.750$ & $25$ & $2000$ \\
$0.44$ & $0.060$ & $8$ & $13$ & $0.500$ & $0.750$ & $25$ & $2000$ \\
$0.44$ & $0.060$ & $10$ & $13$ & $0.519$ & $0.730$ & $19$ & $2000$ \\
$0.44$ & $0.060$ & $12$ & $14$ & $0.519$ & $0.730$ & $19$ & $1300$ \\
\hline
$0.44$ & $0.080$ & $6$ & $12$ & $0.475$ & $0.725$ & $20$ & $2000$ \\
$0.44$ & $0.080$ & $8$ & $13$ & $0.475$ & $0.725$ & $20$ & $2000$ \\
$0.44$ & $0.080$ & $10$ & $13$ & $0.500$ & $0.725$ & $20$ & $1200$ \\
$0.44$ & $0.080$ & $12$ & $14$ & $0.509$ & $0.725$ & $19$ & $1300$ \\
\hline
$0.44$ & $0.100$ & $6$ & $13$ & $0.445$ & $0.725$ & $23$ & $2000$ \\
$0.44$ & $0.100$ & $8$ & $14$ & $0.445$ & $0.725$ & $23$ & $2300$ \\
$0.44$ & $0.100$ & $10$ & $13$ & $0.445$ & $0.725$ & $23$ & $2200$ \\
$0.44$ & $0.100$ & $12$ & $15$ & $0.475$ & $0.725$ & $20$ & $880$ \\
\hline
$0.44$ & $0.120$ & $6$ & $13$ & $0.425$ & $0.725$ & $30$ & $2500$ \\
$0.44$ & $0.120$ & $8$ & $14$ & $0.425$ & $0.725$ & $30$ & $2200$ \\
$0.44$ & $0.120$ & $10$ & $15$ & $0.425$ & $0.725$ & $30$ & $1000$ \\
$0.44$ & $0.120$ & $12$ & $16$ & $0.445$ & $0.725$ & $25$ & $1000$ \\
\hline
$0.44$ & $0.130$ & $6$ & $13$ & $0.375$ & $0.750$ & $20$ & $2000$ \\
$0.44$ & $0.130$ & $8$ & $15$ & $0.375$ & $0.750$ & $20$ & $1000$ \\
$0.44$ & $0.130$ & $10$ & $16$ & $0.375$ & $0.750$ & $20$ & $1000$ \\
\hline
$0.44$ & $0.180$ & $6$ & $14$ & $0.270$ & $0.750$ & $45$ & $3000$ \\
$0.44$ & $0.180$ & $8$ & $16$ & $0.270$ & $0.750$ & $45$ & $1500$ \\
$0.44$ & $0.180$ & $10$ & $19$ & $0.270$ & $0.750$ & $45$ & $512$ \\
\hline
$0.44$ & $0.200$ & $6$ & $14$ & $0.270$ & $0.750$ & $45$ & $3000$ \\
$0.44$ & $0.200$ & $8$ & $16$ & $0.270$ & $0.750$ & $45$ & $2000$ \\
$0.44$ & $0.200$ & $10$ & $19$ & $0.270$ & $0.750$ & $45$ & $512$ \\
\hline
\hline
\end{tabular*}
}
\end{table}

\begin{table}
\caption{
Simulation parameters at finite temperature and $x=0.65$. For details
see Table \ref{tab:simparams032}.
\label{tab:simparams065}}
{\scriptsize
\begin{tabular*}{\columnwidth}{@{\extracolsep{\fill}} l l r c r r c r}
\hline
\hline
$x$ & $h$ & $L$ & $b$  & $T_{\rm min}$ & $T_{\rm max}$ & $N_{T}$ & $N_{\rm sa}$ \\
\hline
$0.65$ & $0.000$ & $6$ & $10$ & $0.500$ & $1.400$ & $20$ & $1000$ \\
$0.65$ & $0.000$ & $8$ & $12$ & $0.500$ & $1.400$ & $20$ & $500$ \\
$0.65$ & $0.000$ & $10$ & $12$ & $0.500$ & $1.400$ & $20$ & $450$ \\
$0.65$ & $0.000$ & $12$ & $11$ & $0.500$ & $1.400$ & $20$ & $500$ \\
$0.65$ & $0.000$ & $14$ & $10$ & $0.850$ & $1.400$ & $15$ & $490$ \\
$0.65$ & $0.000$ & $16$ & $11$ & $0.850$ & $1.400$ & $15$ & $470$ \\
\hline
$0.65$ & $0.050$ & $6$ & $8$ & $0.697$ & $1.400$ & $15$ & $1000$ \\
$0.65$ & $0.050$ & $8$ & $10$ & $0.697$ & $1.400$ & $15$ & $500$ \\
$0.65$ & $0.050$ & $10$ & $9$ & $0.800$ & $1.400$ & $20$ & $500$ \\
$0.65$ & $0.050$ & $12$ & $11$ & $0.697$ & $1.400$ & $15$ & $300$ \\
$0.65$ & $0.050$ & $14$ & $10$ & $0.900$ & $1.400$ & $15$ & $500$ \\
$0.65$ & $0.050$ & $16$ & $11$ & $0.920$ & $1.400$ & $16$ & $280$ \\
\hline
$0.65$ & $0.100$ & $6$ & $8$ & $0.820$ & $1.400$ & $20$ & $750$ \\
$0.65$ & $0.100$ & $8$ & $9$ & $0.820$ & $1.400$ & $20$ & $500$ \\
$0.65$ & $0.100$ & $10$ & $10$ & $0.820$ & $1.400$ & $20$ & $500$ \\
$0.65$ & $0.100$ & $12$ & $11$ & $0.820$ & $1.400$ & $20$ & $800$ \\
$0.65$ & $0.100$ & $14$ & $12$ & $0.820$ & $1.400$ & $20$ & $380$ \\
\hline
$0.65$ & $0.150$ & $6$ & $8$ & $0.820$ & $1.400$ & $20$ & $1000$ \\
$0.65$ & $0.150$ & $8$ & $10$ & $0.820$ & $1.400$ & $20$ & $500$ \\
$0.65$ & $0.150$ & $10$ & $11$ & $0.820$ & $1.400$ & $20$ & $500$ \\
$0.65$ & $0.150$ & $12$ & $12$ & $0.820$ & $1.400$ & $20$ & $500$ \\
$0.65$ & $0.150$ & $14$ & $13$ & $0.820$ & $1.400$ & $20$ & $500$ \\
\hline
$0.65$ & $0.200$ & $6$ & $10$ & $0.661$ & $1.400$ & $15$ & $500$ \\
$0.65$ & $0.200$ & $8$ & $11$ & $0.661$ & $1.400$ & $15$ & $400$ \\
$0.65$ & $0.200$ & $10$ & $14$ & $0.756$ & $1.400$ & $13$ & $1300$ \\
$0.65$ & $0.200$ & $12$ & $15$ & $0.756$ & $1.400$ & $13$ & $1000$ \\
$0.65$ & $0.200$ & $14$ & $16$ & $0.756$ & $1.400$ & $13$ & $990$ \\
\hline
$0.65$ & $0.300$ & $6$ & $10$ & $0.600$ & $1.400$ & $20$ & $750$ \\
$0.65$ & $0.300$ & $8$ & $15$ & $0.600$ & $1.400$ & $20$ & $1000$ \\
$0.65$ & $0.300$ & $10$ & $17$ & $0.600$ & $1.400$ & $20$ & $550$ \\
$0.65$ & $0.300$ & $12$ & $19$ & $0.600$ & $1.400$ & $20$ & $530$ \\
\hline
\hline
\end{tabular*}
}
\end{table}

\vspace*{8cm}

\begin{table}
\caption{
Critical parameters estimated using a finite-size scaling technique:
For each concentration $x$ and field strength $h$ we compute the
critical temperature $T_c$ and critical exponent $\nu$.
\label{tab:numresults}}
{\scriptsize
\begin{tabular*}{\columnwidth}{@{\extracolsep{\fill}} l l l l}
\hline
\hline
$x$ & $h$ & $T_c$ & $\nu$ \\
\hline
$0.32$ & $0.000$ &  $0.340(3)$ & $0.83(3)$ \\
$0.32$ & $0.005$ & $0.335(3)$ & $0.89(3)$ \\
$0.32$ & $0.010$ & $0.340(3)$ & $0.81(3)$ \\
$0.32$ & $0.020$ & $0.314(5)$ & $1.04(7)$ \\
$0.32$ & $0.030$ & $0.281(13)$ & $1.23(17)$ \\
$0.32$ & $0.035$ & $0.269(14)$ & $1.31(19)$ \\
$0.32$ & $0.040$ & $0.247(14)$ & $1.41(15)$ \\
\hline
$0.44$ & $0.000$ & $0.584(1)$ & $0.75(1)$ \\
$0.44$ & $0.020$ & $0.581(1)$ & $0.70(1)$ \\
$0.44$ & $0.040$ & $0.563(3)$ & $0.77(2)$ \\
$0.44$ & $0.060$ & $0.548(5)$ & $0.86(3)$ \\
$0.44$ & $0.080$ & $0.522(5)$ & $0.99(4)$ \\
$0.44$ & $0.100$ & $0.506(5)$ & $1.01(4)$ \\
$0.44$ & $0.120$ & $0.466(9)$ & $1.39(12)$ \\
\hline
$0.65$ & $0.000$ & $0.9597(8)$ & $0.79(2)$ \\
$0.65$ & $0.050$ & $0.9531(10)$ & $0.76(2)$ \\
$0.65$ & $0.100$ & $0.9264(14)$ & $0.84(3)$ \\
$0.65$ & $0.150$ & $0.8832(21)$ & $0.91(4)$ \\
$0.65$ & $0.200$ & $0.8312(41)$ & $1.06(10)$ \\
$0.65$ & $0.300$ & $0.6905(113)$ & $1.11(12)$ \\
\hline
\hline
\end{tabular*}
}
\end{table}
\clearpage

\end{document}